\documentclass[journal]{IEEEtran}

\usepackage{graphicx,color, psfrag}

\ifCLASSINFOpdf
\else
 \fi
\begin{document}
%
\title{Non-negative matrix factorization pansharpening:
 an application to mid-infrared astronomy}

\author{Olivier~Berne,       A. G. G. M.~Tielens, Paolo~Pilleri
        and~Christine~Joblin
\thanks{
Olivier Bern\'e and A. G. G. M. Tielens are with the Leiden Observatory, P.O. Box 9513
NL-2300 RA  Leiden, The Netherlands 

Paolo Pilleri and Chrsitine Joblin are with the Universit\'e de Toulouse ; UPS ; CESR ; 
9 ave colonel Roche, F-31028 Toulouse cedex 9, France, CNRS; UMR 5187; 31028 Toulouse, France

This work is based on observations made with the Spitzer Space Telescope, which is
operated by the Jet Propulsion Laboratory, California Institute of
Technology under NASA contract 1407.

This work has been submitted to the IEEE for possible publication. Copyright may be 
transferred without notice, after which this version may no longer be accessible.

}
}

\markboth{Journal of \LaTeX\ Class Files,~Vol.~6, No.~1, January~2007}%
{Shell \MakeLowercase{\textit{et al.}}: Bare Demo of IEEEtran.cls for Journals}
\maketitle

\begin{abstract}
Mid-infrared  astronomy (operating at wavelengths ranging from 2 to 25 $\mu$m) has 
progressed significantly in the last decades, thanks to the improvement of detector techniques 
and the growing diameter of telescopes. Space observatories benefit from the absence of 
atmospheric absorption, allowing to reach the very high sensitivities needed to perform 3D 
hyperspectral observations, but telescopes are limited in diameter ($< 1$ meter) and therefore
provide observations at low angular resolution (typically a few seconds of arc).
On the other hand, ground-based facilities suffer from strong atmospheric absorption but use large telescopes (above 8m diameter) to perform sub-arcsecond angular resolution imaging through selected windows in the mid-infrared range. In this Paper, we present a method based on Lee and Seung's Non-negative Matrix Factorization (NMF) to merge data from space and ground based mid-infrared (mid-IR) telescopes in order to combine the best sensitivity, spectral coverage and angular resolution. We prove the efficiency of this technique when applied to real mid-IR astronomical data. We suggest that this method can be applied to any combination of low and high spatial resolution positive hyperspectral datasets, as long as the spectral variety of the data allows decomposition into components using NMF.
\end{abstract}

\begin{IEEEkeywords}
Astronomy, Telescopes, Spectroscopy, Signal processing, Hypercubes
\end{IEEEkeywords}

\IEEEpeerreviewmaketitle

\section{Introduction}
\label{intro}

In galaxies (including our own Galaxy, the Milky-Way), the ultraviolet (UV) and visible light  (respectively 
wavelength ranges of 10-400 and 400-750 nm) emitted by stars is absorbed by dust particles having sizes 
ranging from a few nanometers to a few micrometers. The UV-visible energy they absorb
 is then re-emitted in the infrared (IR, 2-1000$\mu$m \cite{dra03}). Therefore, in the recent
years, astronomers have focused their efforts in studying the Milky-Way and external
galaxies to the IR, where the emitted light contains information both on the amount of absorbed 
energy originating from stars and on the composition of interstellar dust. Unfortunately, most of
the IR light coming from space is absorbed by the atmosphere. This has motivated 
a number of IR space missions that provide the sensitivity required to perform hyperspectral 
observations. However, due to technical and cost constrains, space mission can only launch small diameter
telescopes, which limits the spatial resolution of the obtained data. Conversely, the ground-based
 telescopes  working in the IR usually have large apertures providing subarcsecond
angular resolution, but low sensitivities, that only permit imaging through a few broad band filters.
In the field of remote sensing, a set of methods referred to as \emph{pansharpening} \cite{alp07}
have been developed in order to perform data fusion/improvement by 
combining hyperspectral datasets at different spatial resolutions. 
In this paper, we present such a method, developed
to combine ground and space data, to benefit form the advantages of the two techniques. 
This method is based on the decomposition into \emph{components} of hyperspectral data 
obtained by space telescopes, using Non-Negative Matrix Factorization (NMF), 
followed by non-negative least square fitting of ground-based data using these 
\emph{components} (Sect.\ref{for}). We apply this to real data obtained in the mid-IR range
(5-15 $\mu$m) in order to prove the efficiency of the proposed method (Sect.\ref{app}).

\section{Formalism}
\label{for}

\subsection{Decomposition into components with NMF}\label{nmf}
We define the hyperspectral observations of a region of the sky as a 3 dimensional $m_{s}\times n_{s} \times l_{s}$ matrix  $C_{s}(x, y, \lambda)$ where 
$(x,y)$ define the spatial coordinates and $\lambda$ the spectral index. We assume that all the
points in $C_{s}(x,y, \lambda)$ are positive. We call \emph{spectrum} each vector $x_{s}(p_x, p_y,\lambda)$ recorded at a position $(p_x,p_y)$
over the $l_{s}$ wavelength points.  We define a new positive 2D matrix of observations $X_{s}$, the rows of which 
contain the $m_{s} \times n_{s} = k_{s}$, $x_{s}$ \emph{spectra} of $C_{s}$. We now assume that each \emph{spectrum} $x_{s}$
is the result of a linear combination of a limited number $r_{s}$, with $r_{s}<<k_{s}$, of unknown \emph{source
spectra} written $s_s(\lambda)$ so that:

\begin{equation}
\label{specr}
x_s(p_x, p_y, \lambda)= \sum_{i=1\ldots r_s} a^{i}(p_x, p_y) {s^i_s(\lambda)} ,
\end{equation}
where the $a^i(p_x, p_y)$ coefficients are unknown. This can be re-written under the
following matrix product:

\begin{equation}
X_{s}=A_{s}\times S_{s},
\end{equation}
where $A_s$ is the $ k_s \times r_s$ matrix of unknown coefficients of the linear combinations
and $S_s$ is an $r_s \times l_s$ matrix, the rows of which are the \emph{source spectra}. 
This is a typical blind source separation (BSS) problem \cite{car98}, and can be solved
using multiple methods (e.g. \cite{lee01}, \cite{hyv99}, \cite{gri06}). Here, we concentrate on Non-Negative matrix 
factorization \cite{lee01} that is applicable because $A$ and $S$ are positive.
The objective is to find estimations of $A_{s}$ and $S_{s}$, respectively $W_s$ and $H_s$
so that 
\begin{equation}\label{approxg}
X_{s} \approx W_{s}\times H_{s}.
\end{equation}
This is done by adapting the non-negative matrices $W_{s}$ and $H_{s}$
so as to minimise the squared Euclidian distance $\|X_{s}-W_{s}H_{s}\|^{2}$
or the divergence $D(X_{s}|W_{s}H_{s})$, respectively defined as
\begin{equation}
\label{def1}
\|X_{s}-W_{s}H_{s}\|^{2}=\sum_{ij}(X_{s}^{ij}-(W_{s}H_{s})^{ij})^{2},
\end{equation}

and

\begin{eqnarray}
\label{def2}
D(X_{s}|W_{s}H_{s}) & = &  \sum_{ij}(X_{s}^{ij}\log \frac{X_{s}^{ij}}{(W_{s}H_{s})^{ij}}\\
				&     & -X_{s}^{ij}+(W_{s}H_{s})^{ij}),  
\end{eqnarray}

where the exponents $i$ and $j$ respectively refer to the row and column indexes of the matrices.
The algorithm used to achieve the minimisation of the Euclidian distance is based on the iterative
update rule

\begin{eqnarray}
\label{def3}
&H_{s} \leftarrow H_{s} \frac{(^{T}W_{s}X_{s})}{(^{T}W_{s}W_{s}H_{s})},&  \nonumber\\
&W_{s} \leftarrow W_{s}\frac{(X_{s}~^{T}H_{s})}{(W_{s}H_{s}~^{T}H_{s})} &
\end{eqnarray}

and for divergence

\begin{eqnarray}
&H_{s} \leftarrow H_{s} \frac{\sum_{i}W_{s}^{i}X_{s}^{i}/(W_{s}H_{s})^{i}}{\sum_{j}W^{j}},& \nonumber\\
&W_{s} \leftarrow W_{s} \frac{\sum_{i}H_{s}^{i}X_{s}^{i}/(W_{s}H_{s})^{i}}{\sum_{j}H_{s}^{j}} .&
\end{eqnarray}

Euclidian distance and divergence are non increasing under their respective update rules,
so that starting from random $W_{s}$ and $H_{s}$ matrices, the algorithm will converge towards
a minimum for these criteria. This provides the matrix $H_{s}$ containing the $r_s$ estimated 
\emph{source spectra}.

\begin{figure*}
\begin{center}
\includegraphics[width=13cm, angle=90]{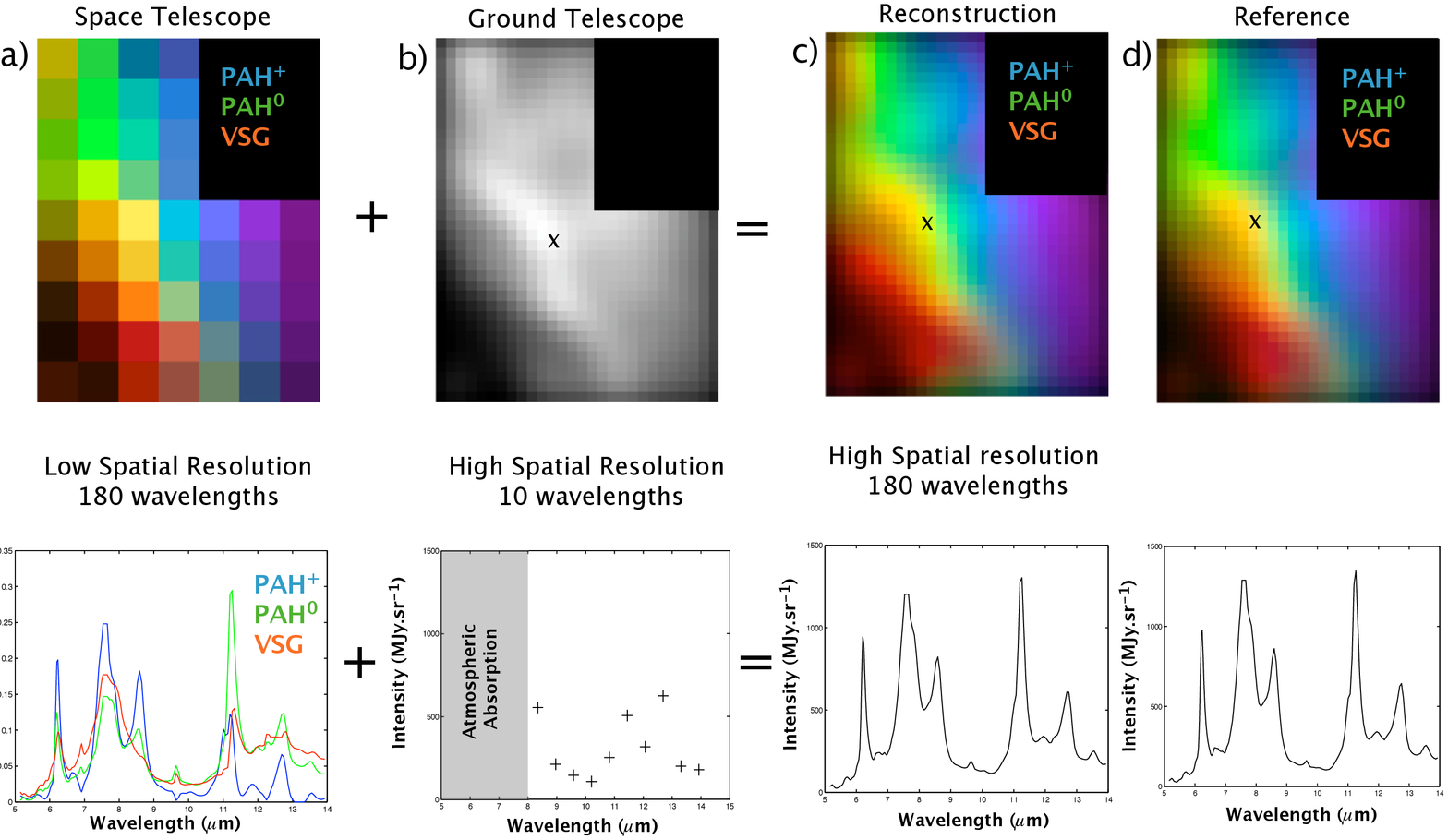}
\caption{ Illustration of the steps achieved to perform pansharpening on mid-IR data of the NGC 7023 nebula: 
{\bf a)} Application of non-negative matrix factorization to the low spatial resolution $C_{~s}$ cube.  
The \emph{source spectra} in $H_{s}$: PAH$^0$, PAH$^+$ and VSG are shown in the lower panel.
The colors in the upper image show the respective contribution, given by $W_s$, of these three 
\emph{source spectra} in the $C_{~s}$ cube at each spatial position.
Colors combine as standard RGB i.e. green+red=yellow etc. {\bf b)}   Upper: image at high 
spatial resolution taken at a given wavelength position from the $C_{~g}$ cube. Lower panel 
shows the spectra at position indicated by a cross in the image. {\bf c)} High spatial/sepctral resolution
$C_{~p}$ cube obtained using NMF pansharpening. 
The contribution of each \emph{source spectrum} from $H_s$ at each spatial position is given with RGB
colors as for the $C_{~s}$ cube. Lower panel shows the reconstructed full mid-IR spectrum
at the position indicated on the image by a cross.
{\bf d)} Reference cube $C_{~ref}$. The colors show the contribution of  each
\emph{source spectra} from $H_s$ to the mid-IR spectrum at each spatial position, obtained by NNLS. Lower panel
shows the reference \emph{spectrum} a the position indicated by a cross in the image.
In all panels, the black mask in the upper right-hand corner of the images is used to mask the contribution 
from a bright star. \label{images}}
\end{center}
\end{figure*}

\subsection{Panshapening}\label{pan}

Once this step has been achieved, one can observe the same region, with a much higher
spatial resolution, but much fewer points in wavelength. This is typically what is done in the 
mid-IR from ground-based telescopes. This will provide a new hyperspectral 
cube $C_{g}(x,y,\lambda)$  with $k_{g}=m_{g}\times n_{g} >> m_{s} \times n_{s} $ spatial positions 
and $l_{g}<<l_{s}$ spectral points. As in Sect. \ref{nmf}, we define an observation matrix $X_{g}$
that contains the \emph{spectra} of $C_{g}$. Again we assume that
 each spectrum in $X_{g}$ can be approximated 
using a linear combination of $r_g$ \emph{source spectra} $s_g$with $l_{g}$ points.
In matrix form, this read:

\begin{equation}\label{approxg}
X_{g} \approx W_{g}\times H_{g}, 
\end{equation}
where $X_{g}$ is the matrix of  spectra observed from the ground, $H_{g}$ is the matrix of \emph{source spectra},
and $W_{g}$ a matrix of unknown coefficients. In fact, the \emph{source spectra} over $l_{s}$ spectral 
points have already been estimated above using NMF and are given by $H_s$.
Because the same region is being observed, the \emph{source spectra} in
$H_g$ are expected to be the same as those in $H_s$ but recorder over
$l_{g}$ spectral positions. Therefore, we construct $H_g$ by taking the
points in $H_s$ corresponding to the $l_g$ spectral positions. 
Now that $X_g$ and $H_g$ are known, Eq. \ref{approxg} no
 longer defines a BSS problem and $W_{g}$
can easily be adjusted by least square methods \cite{law74}. A the end of this process
we therefore have in $W_{g}$ the $k_{g}$ coefficients by which the $H_{g}$ matrix has to be 
multiplied to recover the observed spectrum. Instead, one can build a new \emph{pansharped} matrix 
$X_{p}$ defined by:

\begin{equation}\label{paneq}
X_{p} = W_{g} \times H_{s}.
\end{equation}
This matrix contains the spectra for $k_{g}$ spatial positions and $l_{s}$ wavelength points. 
Finally, the \emph{spectra} in the $X_{p}$ matrix can be 
reassigned their respective positions in an $m_{g}\times n_{g}\times l_{s}$ 
hyperspectral cube $C_{p} (x,y,\lambda)$. The $C_{p}$ cube has therefore
the high number of spatial points of $C_{g}$ and high number of points in 
wavelength of $C_{s}$.

\section{Application to real data}\label{app}

In order to test the above method we have applied it to real astronomical data of the
NGC 7023 nebula. The observed set that will serve as reference (Fig. 1.d) is a spectral cube $C_{ref}$
obtained by the Spitzer space Telescope \cite{wer04} recorded over $36\times 28$ spatial
positions and $180$ points in wavelength in the mid-IR (5-14 $\mu$m). We simulate
a low spatial resolution cube by degrading the spatial resolution of $C_{ref}$ by
a factor of 4. This provides $C_{s}$ that has $9\times7$ spatial points but still
180 points in wavelength (Fig. 1.a). $C_{g}$ is constructed by keeping only 10 points
in wavelengths of $C_{ref}$, corresponding to windows that can be observed from
the ground, but with the full spatial resolution. $C_{g}$ has $36\times 28$ spatial
points but only 10 spectral positions (Fig. 1.b.). The first step consists in applying 
NMF to $C_{s}$ as described in Sect.\ref{nmf}. In order to do this, the
number of rows $r_s$ of $H_{s}$ has to be set. To do this we apply NMF
for successive values of $r_s$ and keep the smallest value that provides:

\begin{equation}
\sum_{i,j} (X_{s}-W_{s}\times H_{s})^2  \sim\sum_{i,j} N_{s}^2
\end{equation}

where $N_{s}$ represents additive contribution of noise to signal in $C_s$. We find
that $r_s=3$ satisfies the above statement, and the 3 obtained spectra of $H_{s}$ are
presented in Fig. 1.a. In the present application, each of the obtained \emph {source spectra}
 can be attributed to the emission of different chemical species: neutral and ionized 
polycyclic aromatic hydrocarbon molecules (resp. PAH$^0$ and PAH$^{+}$)
and very small carbonaceous grains (VSG). This attribution has been discussed in \cite{rap05}
and \cite{ber07} and is beyond the scope of the present paper. However,
we will keep these acronyms in the following for practical reasons.
Using the obtained $H_{s}$ matrix, we construct the $H_{g}$
matrix following the strategy explained in Sect. \ref{pan}. Using the Non Negative Least Square (NNLS) algorithm \cite{law74}, we identify the
values of the coefficients in $W_{g}$. Finally, using equation \ref{paneq} we build $X_{p}$
from $W_{g}$ and $H_{s}$. We then reshape $X_{p}$ in order to recover a spectral
cube $C_{p}$ with $36\times 28$ spatial positions and $180$ points (Fig. 1. c).

\subsection{Efficiency of the method}

The efficiency of the proposed method can be estimated by comparing
$C_{p}$ and $C_{ref}$. Visually, the reconstruction seems very efficient
(See Fig. 1 c))  and d)) for the spatial distribution of the \emph{source
spectra}. Fig. 2 shows an overlay of the original observed spectrum at the position
marked with a cross in Fig. 1 and the reconstructed spectrum. The match
is excellent. Quantitatively, we can estimate the reconstruction efficiency 
of our method, $Q$, defined by:

\begin{equation}
Q=\sqrt{ \frac{ [\sum_{x_i,y_j} \int (C_{p} -C_{ref})^2 d\lambda]}{[\sum_{x_i,y_j} \int C_{ref}^2d\lambda]}},
\end{equation}

which compares the norm of the residuals cube $C_{p} -C_{ref}$ and the norm 
in the reference cube $C_{ref}$. We find that $Q\sim 0.09$ meaning that the
average reconstruction precision is 9$\%$ which is similar to the intrinsic
uncertainty present in the $C_{ref}$ data. The full application presented above 
was implemented under Matlab. Given that this method only involves multiplicative 
calculation, the whole process NMF+Pansharpening runs in less then a 
minute on a laptop computer. We however note that this method will only 
perform for datasets which show significant spectral variations, i.e. providing 
a variety of observed mixtures that is sufficient to achieve NMF. Furthermore, the
number of spectral points obtained in the high spatial resolution dataset must
be sufficient (typically about 10) to allow a unique solution to NNLS fitting.

\subsection{Astrophysical implications}

We have proven the efficiency of the proposed method to obtain
"pansharped"  cubes with full spatial and spectral samplings, combining
data from space and ground-based observatories. This has several
advantages for astronomical observations:

- Astronomical spectra are always positive so this method can easily be applied.

- The method is shown to increase the spatial resolution of mid-IR hyperspecral 
 observations obtained by space telescopes (here by a factor 4).

- The method can provide the full mid-IR spectrum, including in the spectral
region that not accessible from the ground.

- The proposed technique can yield spatially resolved mid-IR
hyperspectral observations of astronomical objects 
for which it would be impossible by classical techniques.

- The procedure involves only multiplicative computations, so is easy to implement
and fast.

The next step consists in obtaining real mid-IR high angular resolution data from a 
ground-based telescope, for astronomical objects previously observed with the 
Spitzer Space Telescope. We have obtained 10 hours of telescope time on the largest single aperture 
telescope in the world, the Grantecan (GTC) telescope, in order to do this. We will then 
be able to combine Spitzer  and GTC observations using non-negative matrix
pansharpening. This will provide the highest angular resolution (0.2 arceseconds) mid-IR
spectral cube ever obtained for an astronomical object.

\begin{figure}
\begin{center}
%
%
\begin{psfrags}%
\psfragscanon%
%
\psfrag{s03}[t][t]{\color[rgb]{0,0,0}\setlength{\tabcolsep}{0pt}\begin{tabular}{c}Wavelength ($\mu$m)\end{tabular}}%
\psfrag{s04}[b][b]{\color[rgb]{0,0,0}\setlength{\tabcolsep}{0pt}\begin{tabular}{c}Intensity (MJy.sr$^{-1}$)\end{tabular}}%
%
\psfrag{x01}[t][t]{6}%
\psfrag{x02}[t][t]{8}%
\psfrag{x03}[t][t]{10}%
\psfrag{x04}[t][t]{12}%
\psfrag{x05}[t][t]{14}%
%
\psfrag{v01}[r][r]{0}%
\psfrag{v02}[r][r]{500}%
\psfrag{v03}[r][r]{1000}%
\psfrag{v04}[r][r]{1500}%
%
\resizebox{8cm}{!}{\includegraphics{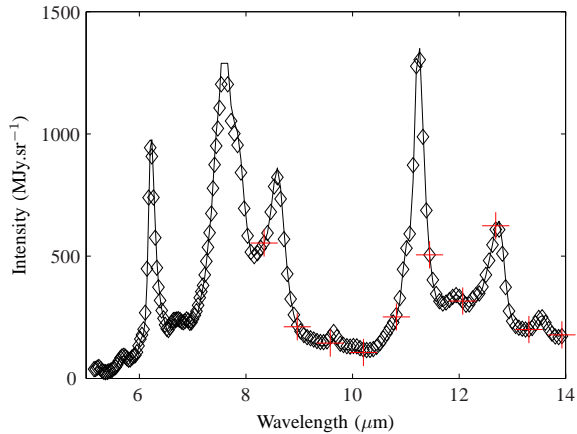}}%
\end{psfrags}%
%

\caption{ A 10 points spectrum taken from $C_g$ at the position
marked with a cross in Fig. 1(red crosses). The continuous line
shows the spectrum taken from $C_p$ at the same position, reconstructed
by fitting the ten points with a linear combination of \emph{source spectra}. 
In diamonds is the reference spectrum taken in $C_{ref}$ at the 
same position. \label{plot}}
\end{center}
\end{figure}

\section{conclusion}

We have proposed an NMF-based pansharpening method that allows to benefit from the high spatial
resolution of mid-IR instruments in ground-based observatories and
sensitivity and spectral coverage of space telescopes. The only working
hypothesis is that the data is positive, which in the case of remote sensing
is usually the case. We have successfully applied this technique to real astronomical
observations. Promising applications could be performed
in the context of future ground- and space-based mid-IR missions.
In particular, the NASA James Webb Space Telescope (JWST hereafter) and
SPICA (JAXA) space missions will provide mid-IR spectral cubes
at angular resolutions of 0.2 arcseconds. Meanwhile, the future
ground-based, 42 meters, European Large Telescope (ELT hereafter) will observe
at milliarcsecond angular resolutions. The combination of
JWST/SPICA data with ELT data using NMF pansharpening
will provide data at the scale size of, for example, the habitable
zone of planet forming disks around young stellar objects. 
Finally, we emphasize the fact that NMF pansharpening 
can well be apply to any combination of low and high
spatial resolution hyperspectral datasets for which an NMF decomposition
into parts is found.

\section*{Acknowledgment}

The authors would like to thank the French PCMI program
for financial support. 

\ifCLASSOPTIONcaptionsoff
  \newpage
\fi

\bibliographystyle{IEEEtran}
\bibliography{refs}



\end{document}